\documentclass[letterpaper, 10 pt, conference]{ieeeconf}

 \IEEEoverridecommandlockouts                              

\usepackage{changes}
\usepackage{cite}
\usepackage{amsmath,amssymb,amsfonts}
\usepackage{algorithm}
\usepackage{comment}
\usepackage{algpseudocode} 
\usepackage{tabularx}
\algtext*{EndWhile}
\algtext*{EndIf}
\algtext*{EndFor}

\usepackage{graphicx}
\usepackage{textcomp}
\usepackage{xcolor}

\usepackage{paralist} 
\usepackage{todonotes}
\usepackage{changes}

\usepackage{tikz}
\usetikzlibrary{shapes,arrows,fit,backgrounds,automata}
\usetikzlibrary{shapes.geometric,fit}
\usepackage[utf8]{inputenc}

\tikzset{main node/.style={circle,draw,minimum size=1cm,inner sep=0pt},}
\tikzstyle{container} = [draw, rectangle, inner sep=0.1cm]
\tikzstyle{container-ellipse} = [draw, ellipse, inner sep=0.1cm]

\usepackage{acronym}
\usepackage{amsthm}

\usepackage{soul}
\usepackage{orcidlink}

\newif\ifuseboldmathops
\newif\ifuseittextabbrevs
\useboldmathopstrue   

\ifuseittextabbrevs

	\newcommand{\ie}{{\it i.e.}}

\else

	\newcommand{\ie}{i.e.~}

\fi

\ifuseboldmathops
	\newcommand{\reals}{\mathbf{R}}

\else
	\newcommand{\reals}{\mathbb{R}}

\fi

\ifuseboldmathops

\else

\fi

\ifuseboldmathops

\else

\fi

\newcommand{\obs}{\mathsf{Obs}}

\newcommand{\dist}[1]{\mathcal{D}(#1)}

\newcommand{\calA}{\mathcal{A}}

\newcommand{\supp}{\mathsf{Supp}}

\newcommand{\plays}{\mathsf{Plays}}

\newcommand{\calAP}{\mathcal{AP}}

\newcommand{\opaqueOBS}{\mathsf{OpqObs}}

\newcommand{\Always}{\Box \, }
\newcommand{\Eventually}{\Diamond \, }
\newcommand{\Next}{\bigcirc \, }

\newcommand{\init}{{\iota}}

\acrodef{mdp}[MDP]{Markov decision process}

\acrodef{asw}[ASW]{Almost-Sure Winning}
\acrodef{ltlf}[LTL$_f$]{Linear Temporal Logic over Finite Traces}
\acrodef{ltl}[LTL]{Linear Temporal Logic}
\acrodef{scltl}[co-safe LTL]{syntaxically co-safe Linear Temporal Logic}
\acrodef{dfa}[DFA]{Deterministic Finite Automaton}

\acrodef{fst}[FST]{Finite State Transducer}
\acrodef{fsa}[FSA]{Finite State Automaton}
\acrodef{nfa}[NFA]{Nondeterministic Finite Automaton}

\newcommand{\Untill}{\mbox{$\, {\sf U}\,$}}

\newcommand{\last}{\mathsf{Last}}

 \newcommand{\hruns}{\mathcal{S}_{H}}
  \newcommand{\lruns}{\mathcal{S}_{T}}

  \newcommand{\fstin}{\mathsf{In}}

  \newcommand{\fstout}{\mathsf{Out}}
  \newcommand{\fstrun}{\mathsf{Run}}

 \definecolor{darkgreen}{rgb}{0,0.5,0}

\newtheorem{problem}{Problem}
\newtheorem{example}{Example}

\newtheorem{lemma}{Lemma}

\newtheorem{definition}{Definition}
\newtheorem{theorem}{Theorem}
\newtheorem{corollary}{Corollary}

\title{Planning with Probabilistic Opacity and Transparency: A Computational  Model of Opaque/Transparent Observations}

\author{Sumukha Udupa\orcidlink{0000-0002-7462-9843} and Jie Fu \orcidlink{0000-0002-4470-2827}
\thanks{S. Udupa and J. Fu   are with the Dept. of Electrical and Computer Engineering, University of Florida, Gainesville, Fl 32611 USA. (e-mail: sudupa, fujie@ufl.edu)}\thanks{
This work was sponsored    by the Army Research Laboratory under Cooperative Agreement Number W911NF-22-2-0233    and   by NSF under grant No. 2144113.}}

\begin{document}
\maketitle

\begin{abstract}
The qualitative opacity of a secret is a security property, which means that a system trajectory satisfying the secret is observation-equivalent to a trajectory violating the secret. 
In this paper, we study how to synthesize a control policy that maximizes the probability of a secret being made opaque against an eavesdropping attacker/observer, while subject to other task performance constraints.
 In contrast to the existing belief-based approach for opacity-enforcement,   
we develop an approach that uses the observation function, the secret, and the model of the dynamical systems to construct a so-called opaque-observations automaton that accepts the exact set of observations that enforce opacity.  Leveraging this opaque-observations automaton, we can reduce the optimal planning in Markov decision processes(MDPs)  for maximizing probabilistic opacity or its dual notion, transparency,   subject to task constraints into a constrained planning problem over an augmented-state MDP.  
Finally, we illustrate the effectiveness of the developed methods in robot motion planning problems with opacity or transparency requirements.
 
\end{abstract}

  \section{Introduction} 

Opacity is a security and privacy property that evaluates whether an observer (intruder) can deduce a system's secret by monitoring its behavior. 
A system is opaque if its secret behavior or private information is made uncertain to an adversarial intruder. Initially introduced for cryptographic protocols by Mazaré \cite{mazare2004using}, opacity has since been explored with various notions depending on the nature of the secret. For instance, state-based opacity ensures that the observer cannot discern if a secret state has been visited, language-based opacity enforces that the observer cannot discern if the trajectory is in a set of secret trajectories\cite{bryans2005modelling,bryans2008opacity,saboori2007notions,saboori2014current,keroglou2018probabilistic,berard2015probabilistic}. 



To set the context, consider a scenario involving an autonomous robot (Player 1 or P1) tasked with routine monitoring in a power plant.  
A sensor network is deployed for daily activity monitoring but can be vulnerable to eavesdropping attacks. Should there be an unauthorized Player 2 (P2) observing the sensor network, P1 must make P2 uncertain if a sequence of waypoints has been visited to enforce location security. Now, the planning question is how can P1 ensure maximal opacity to P2, while fulfilling the routine task beyond a specific threshold? 

In this study, we explore language-based opacity and its dual notion, transparency, in stochastic systems with two players, where one player has imperfect observations. In a stochastic system, probabilistic opacity quantifies the security level of a system \cite{berard2015quantifying} by the probability of generating an opaque trajectory. In competitive settings, achieving maximum probabilistic opacity under task constraints is crucial, while in cooperative scenarios like human-robot interactions, maximizing probabilistic transparency is essential.   

The classical approach for opacity enforcement is to construct a belief-based planning problem (see Related Work) where P1 tracks P2's belief state regarding the current or past state. However, this approach is computationally expensive. 
Instead, we present an approach to construct  a computational model of opaque observations, which are observations obtained by state trajectories that enforce opacity. 
 We use a \ac{fst} to encode the observation function of state trajectories and then the synchronization product between the \ac{fst} and secret language, expressed as a \ac{dfa}, to construct another \ac{dfa} accepting the set of opaque observations. A product between the \ac{mdp}, task \ac{dfa}, and the opaque-observations \ac{dfa} is then computed. We show that the maximally opaque (maximally transparent) control policy under the task performance constraints can be solved as a constrained planning problem in this product. 
\subsection{Related Work}
In the supervisory control of discrete event systems (DESs), researchers have developed opacity-enforcing controllers by modelling a control system  as a \ac{dfa} in the presence of a passive observer \cite{saboori2010verification,saboori2007notions,yingrui2023}. 
Different approaches have been proposed to either synthesize \cite{cassez2012synthesis,wu2014synthesis,xie2021opacity} or verify opacity \cite{lin2011opacity,saboori2008verification,wu2013comparative} in deterministic systems. Saboori \textit{et al.} \cite{saboori2012} synthesized a supervisor to enforce qualitative opacity by restricting system behaviors. In \cite{saboori2011coverage}, the authors explore sensor coverage analysis of mobile agents for various state-based notions of opacity.  
Cassez \textit{et al.}
\cite{cassez2012synthesis} suggest  using a dynamic mask to filter observable events and verifying opacity through solving a 2-player safety game. In \cite{xie2021opacity}, Xie \textit{et al.} proposed a nondeterministic supervisor to prevent the observer, aware of the supervisor's nondeterminism, from determining the secret.  
Besides enforcing opacity, sensor attack strategies to compromise opacity have been investigated in  \cite{yao2022sensor},  where they proposed an information structure that records state estimates for both the supervisor and attacker. In game-theoretic approaches to opacity enforcement, Maubert \textit{et al.} \cite{maubert2011opacity} introduced a game where a player with perfect observations aims to enforce current state opacity against another player with imperfect observations. In our previous work \cite{udupa2023opacity} we explored  qualitative opacity enforcement in a stochastic environment with both the players having partial observation.   

In stochastic DESs, several studies delve into quantifying opacity. Saboori \textit{et al.} \cite{saboori2014current} introduced notions of current state opacity for probabilistic finite automata, along with verification algorithms to verify if the system is probabilistically opaque for  a given threshold. 
Yin \textit{et al.} \cite{yin2019infinite} extended this to the notions of infinite and K-step opacity. Keroglou \textit{et al.} \cite{keroglou2018probabilistic} explored model-based opacity, where the defender conceals the system model using a hidden Markov model. B\'erard \textit{et al.} \cite{berard2015probabilistic} extended language-based opacity to $\omega$-regular properties on \ac{mdp}.  B\'erard \textit{et al.} define the notions of symmetrical opacity, and define probabilistic disclosures. They showed that the quantitative questions relating to opacity become decidable for the class of $\omega$-regular secrets.
Liu \textit{et al.} \cite{liu2024approximate} studied approximate opacity in continuous stochastic control systems, developing a verification method for general continuous-state \ac{mdp}s with finite abstractions. We refer to \cite{jacob2016overview} for an up-to-date and comprehensive review.


 Our proposed approach considers probabilistic opacity/transparency-enforcement in MDPs, distinguishing it from qualitative approaches. In existing stochastic methods, verification commonly involves tracking current state estimates or maintaining a belief state  for opacity enforcement. This work presents an approach to construct a   \ac{dfa} that precisely accepts the set of opaque observations to enforce opacity/transparency without the need to track the observer's beliefs/state estimates. 
 And with that, we enforce opacity  while satisfying a task specification.



\paragraph*{Organization}
In Section \ref{sec:notations}, we introduce preliminaries and frame the problem. Section \ref{sec:solution_approach} presents our computational model. In Section \ref{sec:planning_algorithm}, we use the computational model to solve an optimal planning problem. Section \ref{sec:case_study} illustrates the planning problem's application in a gridworld scenario. Finally, we conclude in Section \ref{sec:conclusion}.

\section{Preliminaries and Problem Formulation}
\label{sec:notations}

\paragraph*{Notations} Let $\Sigma$ be a finite set of symbols, called the alphabet. We denote a set of all $\omega$-regular words as $\Sigma^{\omega}$ obtained by concatenating the elements in $\Sigma$ infinitely many times. A sequence of symbols $w=\sigma_0 \sigma_1 \cdots \sigma_n$ with $\sigma_i \in \Sigma$ for any $0\leq i \leq n$, is a finite word. The set $\Sigma^*$ is the set of all finite words that can be generated with $\Sigma$. The length of a word is denoted by $|w|$. For a word $w = uv$, $u$ is a prefix of $v$ for $u \in \Sigma^\ast$ and $v \in \Sigma^{\omega}$. Let $\reals$ denote the set of real numbers. Given a finite set $Z$, the set of probability distributions over $Z$ is represented as $\dist{Z}$. Given $d \in \dist{Z}$, the support of $d$ is $\supp(d)= \{z\in Z\mid d(z)>0\}$. 
%
\subsection{Markov decision process}
We consider the interaction of P1 with a stochastic environment, monitored by P2 with partial observations, modeled as a terminating \ac{mdp} (without the reward function).


\begin{definition} A probabilistic transition system (\ie, an \ac{mdp} without the reward function) is a tuple 
\[
M=(S, A, P, \mu_0, s_\top, s_\bot, \calAP, O, L)
\] 
where $S$ is a finite set of states, $A$ is a finite set of actions.
$s_\top \in S$ is the \emph{initiating state}, which is a unique $\emph{start}$ state. $s_{\bot} \in S$ is the \emph{terminating state}, which is a unique $\emph{sink}$ state. 
The set $A$ includes a special \emph{initiating action} and a \emph{terminating action}, $a_\top$ and $a_\bot$ respectively. $a_\top$ is only enabled in $s_\top$, while $a_\bot$ is enabled for any $s\in S\setminus \{s_\top, s_\bot\}$.
$P:S \times A\rightarrow \dist{S}$ is a probabilistic transition function, with $P(s'|s,a)$ representing the probability of reaching state $s'$ given action $a$ at state $s$. $\mu_0$ is the initial distribution on states $S\setminus \{s_\top\}$
 such that $\mu_0(s)$ represents the probability of $s$ being the initial state.
For $s_\top \in S$, $P(s_\top, a_\top, s)=\mu_0(s)$. 
That is, if P1 selects the initiating action $a_\top$, then an initial state is reached surely.
For any $s\in S\setminus \{s_\top, s_\bot\}$, $P(s, a_\bot, s_\bot)=1$. That is, if P1 selects the terminating action $a_\bot$, then a terminating state $s_\bot$ is reached surely. Finally, for any $a\in A\setminus\{a_\top, a_\bot\}$, $P(s_\bot,a,s_\bot)=1$, \ie, all allowed actions form a self-loop.
$\calAP$ is a set of atomic propositions, $L: S\rightarrow 2^{\calAP}$ is the labeling function that maps a state to a set of atomic propositions that evaluate true at that state. The initiating state is labeled with a ``starting'' symbol, $\rtimes$, \ie, $L(s_\top) = \rtimes$. The terminating state is labeled the ``ending'' symbol $\ltimes$, \ie, $L(s_\bot) = \ltimes$. $O \subseteq 2^S$ is the set of all finite observables.
 \end{definition}
A finite play $\rho = s_\top a_\top s_0 a_0 s_1\ldots s_n a_\bot s_\bot$ is a sequence of interleaving states and player's actions, where for all $i\ge 0$, $P(s_{i+1}\mid s_i, a_i) > 0$. Let $\plays(M) \subseteq (S\times A)^\ast S$ be the set of finite plays that can be generated from the game $M$. The labeling of a finite play, denoted by $L(\rho)$, is defined as $L(\rho) = \rtimes L(s_0)L(s_1)\ldots L(s_{n})\ltimes$ \ie, the labeling function omits the actions from the play and applies to states only. 


A randomized, finite-memory strategy is a function $\pi: \plays(M) \rightarrow \dist{A}$ mapping a play to a distribution over actions. A randomized, Markov strategy is a function $\pi: S \rightarrow \dist{A}$ mapping the current state to a distribution over actions. The stochastic process generated by applying a policy $\pi$ to the \ac{mdp} $M$ is denoted as $M^{\pi}$, with transition dynamics $P^\pi(s\mid s_\top) = \mu_0(s)$ for the initiating state and $P^\pi(s'\mid s)=\sum_{a\in A} P(s,a,s')\cdot \pi(a\mid s)$ for all $s\in S\setminus \{s_\top\}$. 

 \paragraph*{Information structure} We assume that
\begin{inparaenum}
    \item P1 has perfect observation of states and actions.
    \item P2 has partial observation of states and no observation of P1's actions. 
\end{inparaenum}
 
 \begin{definition}[Transition-Observation function]
 The transition-observation function of P2 is $\obs:S\times A\times S\rightarrow O$ mapping 
a transition $(s,a,s')$ to an observation $o\in O$.
    The observation function of P2 is extended to plays in $M$ as $\obs: \plays(M)\rightarrow O^\ast$ such that for any 
play  $\rho = \rho_1 \cdot s_n a_n s_{n+1}$, $\obs(\rho) = \obs(\rho_1)\cdot \obs(s_n a_n s_{n+1})$ which is the observation of the prefix $\rho_1$ concatenated with the observation of the last transition. 
    Two plays $\rho_1,\rho_2$ are observation-equivalent iff $\obs(\rho_1) = \obs(\rho_2)$. 
\end{definition}

The inverse observation function of P2 $\obs^{-1}: \obs(M) \rightarrow 2^{\plays(M)}$ maps each $\eta \in \obs(M)$ to the set $\obs^{-1}(\eta) = \{\rho \in \plays(M)\mid \obs(\rho) = \eta \}$.
 Given a play $\rho \in \plays(M)$, we denote the set of plays that are observation-equivalent to $\rho$ in P2's perspective by $[\rho]_2$.
 
\textbf{Objective and secret in temporal logic}
In the game arena, P1 aims to achieve a temporal objective $\psi$ specified by a \ac{ltlf} formula, while P2 observes P1. P1 also holds a secret \ac{ltlf} formula $\varphi$ with the goal of concealing whether $\varphi$ is satisfied or not from P2. The two formulas $\psi$ and $\varphi$ may be different.
 
The syntax of the \ac{ltlf} formula is as follows. 

\begin{definition}[\ac{ltlf}~\cite{de2013linear}]
 An (\ac{ltlf}) formula over $\calAP$ is defined inductively as follows:
\[ \varphi := p \mid \neg\varphi \mid \varphi_1 \land \varphi_2 \mid \varphi_1 \lor \varphi_2 \mid \Next \varphi \mid \varphi_1 \Untill \varphi_2 \mid \Eventually \varphi \mid \Always \varphi, \] 
where $p \in \calAP$; $\neg$, $\land$ and $\lor$ are the Boolean operators: negation, conjunction and disjunction, respectively; and $\Next$, $\Untill$, $\Eventually$ and $\Always$ denote the temporal modal operators for \texttt{next}, \texttt{until}, \texttt{eventually} and \texttt{always}, respectively.
\end{definition}

The operator $\Next \varphi$ specifies that formula $\varphi$ holds at the next time instant. $\varphi_1 \Untill \varphi_2$ denotes that there exists a future time instant at which $\varphi_2$ holds and that $\varphi_1$ holds at all time instants up to and including that future instant. The formula $\Eventually \varphi$ specifies that $\varphi$ holds at some future instant, and $\Always \varphi$ specifies that $\varphi$ holds at all current and future time instants.
See~\cite{de2013linear} for detailed semantics of \ac{ltlf}.

For any \ac{ltlf} formula $\varphi$ over $AP$, a set of words $\mathsf{Words}(\varphi) =\{w \subseteq (2^{AP})^* \mid w\models \varphi\}$ that satisfy the formula is associated. 

\begin{definition}[\ac{dfa}]
A \ac{dfa} is a tuple $\mathcal{A}=(Q,\Sigma,\delta,\init, F)$ with a finite set of states $Q$, a finite alphabet $\Sigma$, a deterministic transition function $\delta:Q\times \Sigma \rightarrow Q$, extended recursively as $\delta(q, \sigma u)=\delta(\delta(q,\sigma),u)$ for $\sigma \in \Sigma, u \in \Sigma^{\ast}$.  
$\init$ is the initial state, and $F\subseteq Q$ is a set of accepting states. 
\end{definition}

 A \ac{nfa} is defined similarly to a \ac{dfa}, with a nondeterministic transition function $\delta:Q\times \Sigma \rightarrow 2^{Q}$, where for every $q\in Q$ and $\sigma\in \Sigma$, a transition occurs to a set of states $X \in 2^Q$. 

  Using the method of De Giacomo and Vardi \cite{de2013linear}, we convert the \ac{ltlf} formulae into \ac{dfa} that accepts $\mathsf{Words}(\varphi)$ with $\Sigma = 2^{\mathcal{AP}}$. 
 We have the \ac{dfa} accepting all the words satisfying the secret $\varphi$ as $\calA_s = (Q_s, 2^{\mathcal{AP}}, \delta_s, \init_s, F_s)$, and the \ac{dfa} accepting all words satisfying the task specification $\psi$ as $\calA$. Both \ac{dfa}s are considered to be complete \footnote{An incomplete \ac{dfa} can be completed by adding a sink state and redirecting all undefined transitions to that sink state.}.


\subsection{Probabilistic opacity and transparency}
We introduce symmetric opacity as in \cite{berard2015quantifying}.
\begin{definition}[Symmetrical Opacity]
   An \ac{ltlf} formula $\varphi$ is opaque to P2 with respect to a play $\rho \in \plays(M)$ iff either of the following cases hold:  \begin{inparaenum}
       \item $L(\rho)\models \varphi$; and 
there exists at least one observation-equivalent play $\rho' \in [\rho]_2 $ with $L(\rho') \not\models \varphi$; or
\item $L(\rho)\not\models \varphi$; and there exists at least one observation-equivalent play $\rho' \in [\rho]_2$ with $L(\rho') \models \varphi$.
   \end{inparaenum}
\end{definition}

We now define transparency as a dual property of opacity.
\begin{definition}[Symmetrical Transparency] 
An \ac{ltlf} formula $\varphi$ is transparent to P2 with respect to a play $\rho \in \plays(M)$ iff either of the following cases hold:   \begin{inparaenum}
       \item $L(\rho)\models \varphi$; and for any observation-equivalent play $\rho' \in [\rho]_2$,  $L(\rho')  \models \varphi$, or
\item $L(\rho)\not\models \varphi$; and for any observation-equivalent play $\rho' \in [\rho]_2$ with $L(\rho') 
\not \models \varphi$.
   \end{inparaenum}
    
\end{definition}

We define the set of opaque and non-opaque runs for the secret $\varphi$ and observation function $\obs$ as follows. 

\begin{itemize}
    \item  The set of opaque runs is as given below.
\begin{multline}
\hruns = \{\rho \in \plays(M) \mid \exists \rho' \in [\rho]_2, (L(\rho) \not \models \varphi \land \\ L(\rho') \models \varphi) \lor  (L(\rho) \models \varphi \land L(\rho') \not\models \varphi)\} 
\end{multline}

    


\item   The set of transparent   runs is  
 \begin{multline}
  \lruns = \{\rho \in \plays(M) \mid \forall  \rho' \in  [\rho]_2, \\
  (L(\rho) \models \varphi  \implies L(\rho') \models \varphi)\\
  \lor (L(\rho) \not \models \varphi \implies L(\rho') \not\models \varphi) \} 
\end{multline}


    
\end{itemize}

It is easy to see $\lruns \cap \hruns = \emptyset$. 
\begin{lemma}
\label{lemma: opaque_union_transparent_is_plays}
    $\hruns \cup \lruns = \plays(M)$.
\end{lemma}
The proof follows from the definition and thus is omitted.
The following definition of quantitative opacity has been introduced in  \cite{berard2015probabilistic} for probabilistic systems. 
    \begin{definition}[Probabilistic opacity and transparency] 
    Given the \ac{mdp} $M$, observation function $\obs$, the secret $\varphi$ and policy $\pi$, the probabilistic opacity of $\varphi$ in the stochastic process $M^\pi$ is 
    \[
     PH(\varphi; M^\pi, \obs) = \Pr(\hruns; M^\pi  )
    \]
 where $\Pr(\hruns;M^\pi)$ denotes the probability of an opaque run given $M^\pi$. 
 The probabilistic transparency of $\varphi$ in $M^{\pi}$ is  
        \[
    PT (\varphi; M^\pi, \obs) = \Pr(\lruns; M^\pi  )
    \] 
\end{definition}

We also define the set of runs that satisfy the task specification as given below.
\begin{equation}
    \mathcal{S}_{\psi} = \{\rho \in \plays(M) \mid L(\rho) \models \psi\}
\end{equation}
 
 \begin{problem}
 \label{prob:problem_statement}
 Given the \ac{mdp} $M$, secret $\varphi$, task $\psi$ and the observation function $\obs$, compute a strategy $\pi^\ast$ for P1 that maximizes the probabilistic opacity of $\varphi$ while ensuring that the probability of satisfying the task $\psi$ is greater than $\epsilon$.
Formally, \begin{equation}
    \begin{aligned}
         \max PH(\varphi; M^{\pi}, \obs) \\ 
        \textrm{s.t.} \Pr(\mathcal{S}_{\psi}; M^{\pi}) \geq \epsilon.
    \end{aligned}
\end{equation}
 \end{problem} 
By replacing the maximization with minimization, and Lemma~\ref{lemma: opaque_union_transparent_is_plays}, the solution to the optimization problem is a policy that maximizes the probabilistic transparency of $\varphi$ under a task performance constraint.

To illustrate our definitions, we use a running example.
 


\begin{figure}[h]
    \centering   
   \vspace{1.5mm} 
   \begin{tikzpicture}[->,>=stealth',shorten >=1pt,auto,node distance=2.5cm,scale=0.7,semithick, transform shape]
    \tikzstyle{every state}=[fill=black!10!white];


    \node[state, initial] (1) at (-3, -0.5) {$s_1$};
    \node[state] (2) at (0, 0.5) {$s_2$};
    \node[state] (3) at (0, -2) {$s_3$};
    \node[state] (4) at (5, 2) {$s_4$};
    \node[state] (5) at (5, 0.5) {$s_5$};
    \node[state] (6) at (5, -2) {$s_6$};
    \node[state] (7) at (4, -3.25) {$s_7$};

    \node [container,fit=(2) (3),draw=red,dashed,line width=0.2mm ] (container) {};
    \node [container,fit=(4),draw=green,dashed,line width=0.2mm ] (container) {};
    \node [container,fit=(5) (6),draw=blue,dashed,line width=0.2mm ] (container) {};
    \node [container,fit=(7),draw=violet,dashed,line width=0.2mm ] (container) {};

      \path 
        (1) edge   node[pos=0.4, sloped]{$a,b (0.5)$} (2)
        (1) edge   node[below, pos=0.4, sloped]{$a, b (0.5)$} (3)
        (2) edge [loop above] node {$a (0.2)$} (2)
        (2) edge node [sloped] {$a (0.8)$} (3)
        (2) edge   node[pos=0.5, sloped]{$b (0.4)$} (4)
        (2) edge   node[pos=0.5]{$b (0.6)$} (5)
        (3) edge [loop below] node {$a (0.2)$} (3)
        (3) edge   node[ pos=0.3,sloped]{$b (0.3)$} (4)
        (3) edge   node[pos=0.6, sloped]{$a (0.4)$, $b (0.2)$} (5)
        (3) edge node[]{$b (0.5)$} (6)
        (3) edge node[sloped]{$a (0.4)$} (7) 
        (4) edge[loop right]   node{$a, b$} (4)
        (5) edge[loop right]   node{$a (0.3)$, $b (1)$} (5)
        (5) edge node[right]{$a (0.2)$} (4)
        (5) edge [out=335, in=335, pos=0.2] node{$a (0.5)$} (7)
        (6) edge [loop right]node{$a (0.5)$, $b (1)$} (6)
        (6) edge node[sloped]{$a (0.5)$} (5)
        (7) edge node[]{$a,b (0.5)$} (6)
        (7) edge [loop left]node{$a, b (0.5)$} (7) 
        ;

\end{tikzpicture}
    \vspace{-13mm}
    \caption{\ac{mdp} for the illustrative example. The colored boxes represent the observation partition of the state space.
    }
    \vspace{-4mm}
    \label{fig:running_example}
\end{figure}
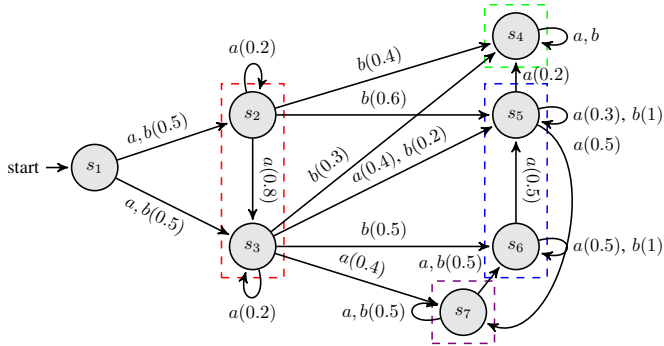

 \begin{example}[Part I]
 \label{example:part_I}
 
    Consider the \ac{mdp} in Fig.\ref{fig:running_example} with $7$ states, $s_1$ through $s_7$ ($s_1$ as the initial state.) Here, $s_4$ is a sink, and edges are labeled with actions and transition probabilities. The transition-observation of P2, $\obs(s,a,s')$, is a set of states that are observation equivalent to $s'$ (actions are not observable). The \ac{mdp} has observation-equivalent states sets $\{s_1\}$, $\{s_2, s_3\}$, $\{s_4\}$, $\{s_5, s_6\}$ and $\{s_7\}$.  
     P1's task is to eventually reach $s_4$ ($\psi = \Eventually s_4$), and secret  formula is satisfied if P1 reaches the state $s_6$, ($\varphi = \Eventually s_6$). P1 must compute a strategy for maximal 
     probabilistic opacity of $\varphi$ while ensuring to satisfy $\psi$ with a probability $\ge \epsilon$. We can also consider the planning problem where P1 aims to maximize the probabilistic transparency with respect to $\varphi$ under the task performance constraint.  
 \end{example}

 \section{Main results}
 \label{sec:solution_approach}
In this section, we present a computational model to obtain the opaque observations of P2. Using this computational model, we formulate constrained \ac{mdp}s to solve Problem~\ref{prob:problem_statement}.
 \subsection{A computational model for  opaque observations}
 

 To solve Problem~\ref{prob:problem_statement}, we need to construct the set $\hruns$ of opaque runs, which is measurable. 
 However, the set of possible runs in $M$ can be infinite, such an enumeration procedure cannot terminate and therefore we cannot construct $\hruns$ for planning purpose. 
We present a method to construct a \ac{dfa} that accepts the set of opaque observations. 
\begin{definition}[Opaque observations]
Given the secret $\varphi$, 
an observation $\rho_o\in O^\ast$ is opaque iff there is an opaque run $\rho\in \hruns$ and $\rho_o  = \obs(\rho)$.     The set of \emph{opaque observations} is $\opaqueOBS$.
\end{definition}

 
First, it is seen that the observation function $\obs: \plays(M)\rightarrow O^\ast$ can be expressed as an \ac{fst} whose input strings are $\plays(M)$ and output strings are the observations of $\plays(M)$.

\begin{definition}[Finite-state transducer representation for the observation function]
The \ac{fst} encoding of the observation function $\obs: \plays(M)\rightarrow O^\ast$ is a tuple 
\[
\mathcal{A}_o =\langle  S, \Sigma_I, \Sigma_O, T_o \rangle 
\]
  where
  \begin{itemize}
      \item $S$ is the finite set of states of the \ac{mdp}.
      \item $ \Sigma_I =  S\times A\times S$ is the set of input symbols, which are the transitions in the \ac{mdp}.  
      \item $\Sigma_O = O \cup \{\rtimes, \ltimes\}$ is the set of output symbols, which are the observations in the \ac{mdp} and two symbols $\rtimes$ and $\ltimes$ marking the beginning and ending of a word. 

      \item $T_o: S\times \Sigma_I\rightarrow S \times \Sigma_O $ is a deterministic transition function that maps a given pair of states and inputs to a next state and an output. It is constructed as follows: 
  \begin{itemize}
      \item if $s_\top \in S$ is the initiating state, the input $(s_{\top} ,a_\top,s)$ is enabled if $\mu_0(s)>0$ in the \ac{mdp} $M$, then
  \[
  T_o(s_{\top},(s_{\top},a_\top,  s ))=(s, \rtimes),
  \]
  \item if $s \in S$, then the input $(s,a,s')$ is enabled if $P(s,a,s')>0$ in the \ac{mdp} $M$,
  \[
    T_o(s,(s, a, s' ) )=(s',o),
  \]
  where $o = \obs(s,a,s')$.
  

 \item for any state $s \in S$, if $a_\bot$ action is enabled, then
  \[
  T_o(s,(s, a_\bot, s_\bot))=(s_\bot, \ltimes).
  \]
  \end{itemize} 

  \end{itemize}
\end{definition}

A transition from state $t$ to $t'$ given input $\sigma \in \Sigma_I$ and output $o\in \Sigma_o$ is also denoted $t\xrightarrow[o]{\sigma}t'$. Given a sequence of inputs, also called input word $w = \sigma_0\sigma_1\ldots \sigma_n$, the transducer generates 
a sequence of transitions, or a run $\rho = t_0 \xrightarrow[o_0]{\sigma_0} t_1 \xrightarrow[o_1]{\sigma_1} t_2\ldots \xrightarrow[o_{n-1}]{\sigma_{n-1}}t_n$ with the output uniquely determined as $w_o = o_0o_1\ldots o_n$.

 By construction, it is noted that for any play $\rho = s_\top a_\top s_0 \ldots s_n a_\bot s_\bot$,  the corresponding input to the \ac{fst} is 

\vspace{-6mm}
\begin{multline*}
    \fstin(\rho) =(s_{\top},a_\top,s_0) (s_0,a_0,s_1)(s_1,a_1,s_2)\ldots \\(s_{n-1},a_{n-1},s_n) (s_n,a_\bot,s_{\bot}). 
\end{multline*}

The output  is 
\begin{multline*}
   \fstout(\rho) = \rtimes \obs(s_0,a_0,s_1) \obs(s_1,a_1,s_2)\ldots\\ \obs(s_{n-1},a_{n-1},s_n)\ltimes . 
\end{multline*}
 The corresponding run is $\fstrun(\rho) = s_{\top} s_0s_1s_2\ldots s_ns_{\bot}.$

\begin{figure}
    \vspace{1mm}
    \centering
    \begin{tikzpicture}[->,>=stealth',shorten >=1pt,auto,node distance=2.5cm,scale=0.7,semithick, transform shape]
    \tikzstyle{every state}=[fill=black!10!white];


    \node[state, initial] (1) at (-3, 3.5) {$s_\top$};
    \node[state] (2) at (0, 3.5) {$s_1$};
    \node[state] (3) at (3.5, 1.5) {$s_2$};
    \node[state] (4) at (-3.5, 1.5) {$s_3$};
    \node[state] (5) at (2.5, -1.5) {$s_4$};
    \node[state] (6) at (-0.3, -0.75) {$s_5$};
    \node[state, accepting] (9) at (-4, -1.5) {$s_\bot$};


      \path 
        (1) edge   node[pos=0.48, sloped]{$(s_\top,a_\top,s_1)\backslash [\rtimes]$} (2)
        (2) edge   node[below, pos=0.4, sloped]{$(s_1, a, s_2)\backslash [s_2, s_3]$} (3)
        (2) edge node [sloped] {$(s_1, b, s_2)\backslash [s_2, s_3]$} (3)
        (2) edge   node[below, pos=0.4, sloped]{$(s_1, b, s_3)\backslash [s_2, s_3]$} (4)
        (2) edge node [sloped] {$(s_1, a, s_3)\backslash [s_2, s_3]$} (4)
        (3) edge [loop right] node [sloped]{$(s_2, a, s_2)\backslash [s_2, s_3]$} (3)
        (3) edge   node[pos=0.5, sloped]{$(s_2, a, s_3)\backslash [s_2, s_3]$} (4)
        (3) edge   node[pos=0.6, sloped]{$(s_2, b, s_5)\backslash [s_5, s_6]$} (6)
        (3) edge   node[pos=0.5, sloped]{$(s_2, b, s_4)\backslash [s_4]$} (5)
        (4) edge [loop above] node[sloped, pos=0.5] {$(s_3, a, s_3)\backslash [s_2, s_3]$} (4)
        (4) edge   node[sloped]{$(s_3, b, s_5)\backslash [s_5,s_6]$} (6)
        (4) edge node[below,sloped]{$(s_3, a, s_5)\backslash[s_5,s_6]$} (6)
        (6) edge node[sloped]{$(s_5,a,s_4)\backslash[s_4]$} (5) 
        (5) edge[loop right]   node[sloped, pos=0]{$(s_4,a,s_4)\backslash[s_4]$} (5)
        (5) edge [pos=0.75]node[sloped]{$(s_4,a_\bot,s_\bot)\backslash[\ltimes]$} (9)
        (4) edge [] node[sloped]{$(s_3,a_\bot,s_\bot)\backslash [\ltimes]$} (9)
        ;

\end{tikzpicture}
    \vspace{-12mm}
    \caption{A fragment of \ac{fst} encoding the observation function.}
    \vspace{-7mm}\label{fig:running_example_fst}
\end{figure}
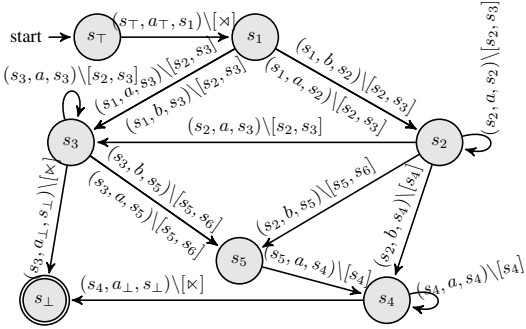

\begin{example}[Part II]
    \label{example:part_II}
    Continuing Example \ref{example:part_I},  a fraction of \ac{fst} encoding of the observation function is shown in Fig.~\ref{fig:running_example_fst}.  Each of the states in the transducer represents the states in the \ac{mdp} shown in Fig.~\ref{fig:running_example} along with initiating ($s_\top$) and terminating ($s_\bot$) states. Transitions are labeled with input and output symbols. For instance, the transition from $s_1$ to $s_2$ is labeled $(s_1,a,s_2)$ as input and $[s_2,s_3]$ as output symbols. 
    Each state also has a transition to the terminating state $s_\bot$. 
\end{example}

The next step is to compute a product between the \ac{fst} of the observation function and the \ac{dfa} accepting the secret.
\begin{definition}[Product \ac{fst}]
Given the \ac{fst} representation of the observation function $\mathcal{A}_o$, and the secret \ac{dfa} $\calA_s$, the product of the \ac{fst} and \ac{dfa} is a tuple, 
    \[
\calA_H = \langle Q_h, \Sigma_I, \Sigma_O, T_h, \init_h, F_h\rangle 
\]
\begin{itemize}
    \item $Q_h = S\times Q_s$ is the set of states.
    \item $\init_h = (s_\top, \init_s)$ is the initial state.
    \item $T_h:Q_h\times \Sigma_I \rightarrow Q_h \times \Sigma_O$ is the transition function and is defined as follows:
    For $\sigma \in S\times A\times S$, if $s \xrightarrow[o]{\sigma} s'$ in $\calA_o$ and $q\xrightarrow[]{a} q'$ in $\calA_s$, then, 
    \begin{enumerate}
\item For $\sigma = (s_\top, a_\top, s)$, and $s_\top \xrightarrow[\rtimes]{\sigma} s$, let 
\[
(s_\top, \init_s) \xrightarrow[\rtimes]{\sigma} (s, q)
\]
where $q= \delta(\init_s, L(s))$. 


\item For $\sigma = (s, a_\bot, s_\bot)$, and $s \xrightarrow[\ltimes]{\sigma} s_\bot$, let 
\[
(s, q) \xrightarrow[\ltimes]{\sigma} (s_\bot, q).
\]



\item For any other $\sigma$, let 
\[
(s,q)\xrightarrow[o]{\sigma}(s', q') \text{ if and only if } \delta_s(q, L(s')) = q'.
\]

\item Any state $(s,q)$ where $s= s_\bot$ is a sink state.
\end{enumerate}
\item $F_h = \{(s_\bot, q)\mid q\in F_s\}$ is a set of accepting states.

\end{itemize}
\end{definition}

From the above construction, we establish a mapping $\mathfrak{R}: \plays(M) \rightarrow \fstrun(\mathcal{A}_H)$ between plays in \ac{mdp} $M$ and runs in the product of the \ac{fst} and \ac{dfa} $\mathcal{A}_H$ as follows: For a play $\rho = s_\top a_\top s_0\ldots s_n a_\bot s_\bot \in \plays(M)$, there is a run (\ie, a state sequence augmented with automata states) $\hat \rho $ such that $\hat \rho = (s_\top, \init_s)(s_0, q_0)\ldots (s_n,q_n)(s_\bot, q_n)$ where $q_i=\delta(q_{i-1}, L(s_i))$ for $i\ge 1$ and $q_0 = \delta(\init_s, L(s_0))$. 



\begin{lemma}
\label{lemma:plays_to_AH_satisfy}
A run $\rho \in \plays(M)$ is such that $L(\rho)\models \varphi$ \\ iff $\mathfrak{R}(\rho)$ ends in $(s_\bot, q)$ where $q\in F_s$.
\end{lemma}
\begin{proof}
By construction, consider a run $\rho = s_\top a_\top \ldots a_\bot s_\bot\\ \in \plays(M)$, if $\mathfrak{R}(\rho)$ ends in $(s_\bot, q)$ with $q\in F_s$, then $L(s_\top s_0\ldots s_ns_\bot) $ is accepted by $\calA_s$. Thus, $L(\rho) \models \varphi$. 
\end{proof}

We denote the last/final state of $\mathfrak{R}(\rho)$ as $\last(\rho)$. We then compute two \ac{nfa}s from the product transducer.

\begin{definition} 
\label{def:NFA_construction}
Given the product \ac{fst} $\calA_H= \langle Q_h, \Sigma_I, \Sigma_O, T_h, \init_h, F_h\rangle $,  the \emph{output language} accepted by $\calA_H$ is obtained as the set $L_O(\calA_H) =\{\fstout(\rho) \mid \rho \in \Sigma_I^\ast, \last(\rho) \in \{s_\bot\}\times F_s \}$.  

The \ac{nfa} accepting $L_O(\calA_H)$ is obtained from $\calA_H$ by removing the input symbols and use the output symbols on each transition as the input symbol. 

\end{definition}

\begin{lemma}
    For any $\rho \in \plays(M)$, if $\rho \models \varphi$, $\fstout(\rho) \in L_O(\calA_H)$ and if $\rho \not\models \varphi$, $\fstout(\rho) \notin L_O(\calA_H)$. 
\end{lemma}
\begin{proof}
    By construction, $\last(\rho) = (s_\bot, q)$ where $q \in F_s$. From Lemma \ref{lemma:plays_to_AH_satisfy}, we have that $L(\rho) \models \varphi$.
\end{proof}
 Also, let $\calA_H^\dagger = \langle Q_h, \Sigma_I, \Sigma_O, T_h, \init_h,  F_h^\dagger \rangle$ where $F_h^\dagger = \{(s_\bot, q)\mid q\in Q_s\setminus F_s\}$. The output language $L_O(\calA_H^\dagger)$ and its \ac{nfa} is obtained  in a similar way.
  
\begin{theorem}
\label{lemma:opaque_runs}
The set of \emph{opaque runs}   is $Z= \obs^{-1}(L_O(\calA_H)\cap L_O(\calA_H^\dagger))$. 
\end{theorem}
\begin{proof}
    From Def. \ref{def:NFA_construction}, for any $u \in L_O(\calA_H)$, there exists an input string $\rho'$ such that $L(\rho' ) \models \varphi$ and $\fstout(\rho')=u $.  For any $u \in L_O(\calA^\dagger_H)$, there exists an input string $\rho''$ such that $L(\rho'' )  \models \neg \varphi$ and $\fstout(\rho')=u $.  Thus, if $\obs(\rho)  = u \in L_O(\calA_H) \cap  L_O(\calA^\dagger_H)$, then $\rho, \rho',\rho''$ are observation-equivalent. On observing $\obs(\rho)$ ($=\obs(\rho')=\obs(\rho'')$), P2 cannot know if $L(\rho)\models \varphi$ or not.

 Next, we aim to show that any run not in $Z$ must be transparent. For any $\rho \notin Z$, $\obs(\rho) \notin L_O(\calA_H)\cap L_O(\calA_H^\dagger)$.  
 Since $L_O(\calA_H)\cup L_O(\calA_H^\dagger) = \obs(\plays(M))$ includes all possible observations, it is  either (I) $\obs(\rho) \in (L_O(\calA_H) \setminus L_O(\calA_H^\dagger) )$ or (II) $\obs(\rho) \in (L_O(\calA^\dagger_H) \setminus L_O(\calA_H ) )$. In Case (I), the observation informs that $\varphi$ is satisfied by $\rho$ because any run violating $\varphi$ is not observation-equivalent to $\rho$. In case (II), the observation informs P2 that $\varphi$ is   violated because any run satisfying $\varphi$ is not observation-equivalent to $\rho$.

\end{proof}

\begin{figure}[ht!]
  \vspace{-10mm}
  \centering
  \begin{minipage}{0.45\linewidth}
    \centering



\begin{tikzpicture}[->,>=stealth',shorten >=1pt,auto,node distance=2.5cm,scale=0.7,semithick, transform shape]

    \tikzstyle{every state}=[fill=black!10!white];
    \node[state, initial] (1) at (-3, -0.5) {$p_1$};
    \node[state, accepting] (2) at (-1.5, -0.5) {$p_2$};
      \path 
        (1) edge [loop above]  node{$\neg s_4$} (1)
        (1) edge   node {$s_4$} (2)
        (2) edge [loop above]   node{$True$} (2)
        ;
\end{tikzpicture}

  \end{minipage}
\hfill
\begin{minipage}{0.45\linewidth}
    \centering



\begin{tikzpicture}[->,>=stealth',shorten >=1pt,auto,node distance=2.5cm,scale=0.7,semithick, transform shape]

    \tikzstyle{every state}=[fill=black!10!white];
    \node[state, initial] (1) at (-3, -0.5) {$q_1$};
    \node[state, accepting] (2) at (-1.5, -0.5) {$q_2$};
      \path 
        (1) edge [loop above]  node{$\neg s_6$} (1)
        (1) edge   node {$s_6$} (2)
        (2) edge [loop above]   node{$True$} (2)
        ;
\end{tikzpicture}

  \end{minipage}
  \vspace{-8mm}
  \caption{
    \textbf{(a)} \ac{dfa} for the task specification $\psi = \Eventually s_4$.
    \textbf{(b)} \ac{dfa} for the secret specification $\varphi = \Eventually s_6$.}
  \label{fig:running_example_dfa}
\end{figure}
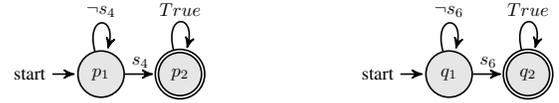

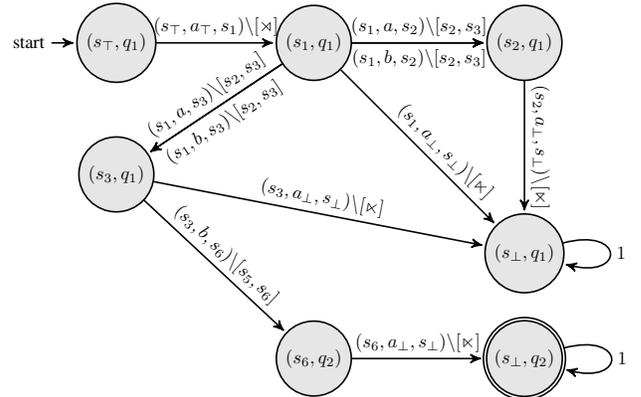
\begin{figure}
    \vspace{-3mm}
    \centering
    \begin{tikzpicture}[->,>=stealth',shorten >=1pt,auto,node distance=2.5cm,scale=0.7,semithick, transform shape]
    \tikzstyle{every state}=[fill=black!10!white];


    \node[state, initial] (0) at (-4.25, 3.5) {$(s_\top,q_1)$};
    \node[state] (1) at (-0.5, 3.5) {$(s_1,q_1)$};
    \node[state] (2) at (3.5, 3.5) {$(s_2,q_1)$};
    \node[state] (3) at (-4.25, 1) {$(s_3,q_1)$};
    \node[state] (5) at (-0.5, -2.5) {$(s_6,q_2)$};
    \node[state] (6) at (3.5, -0.5) {$(s_\bot,q_1)$};
    \node[state, accepting] (7) at (3.5, -2.5) {$(s_\bot, q_2)$};


      \path 
        (0) edge   node[sloped]{$(s_\top,a_\top,s_1)\backslash[\rtimes]$} (1)
        (1) edge   node[sloped]{$(s_1,a,s_2)\backslash[s_2, s_3]$} (2)
        (1) edge   node[below,sloped]{$(s_1,b,s_2)\backslash[s_2, s_3]$} (2)
        (1) edge   node[pos=0.6,sloped]{$(s_1,a_\bot,s_\bot)\backslash[\ltimes]$} (6)
        (1) edge   node[sloped]{$(s_1,a,s_3)\backslash[s_2, s_3]$} (3)
        (1) edge   node[below,sloped]{$(s_1,b,s_3)\backslash[s_2, s_3]$} (3)
        (2) edge   node[sloped]{$(s_2,a_\bot,s_\bot)\backslash[\ltimes]$} (6)
        (3) edge   node[sloped]{$(s_3,a_\bot,s_\bot)\backslash[\ltimes]$} (6)
        (3) edge   node[sloped]{$(s_3,b,s_6)\backslash[s_5, s_6]$} (5)
        (5) edge   node[sloped]{$(s_6,a_\bot,s_\bot)\backslash[\ltimes]$} (7)
        (7) edge [loop right]  node[]{$1$} (7)
        (6) edge [loop right]  node[]{$1$} (6)

        ;

\end{tikzpicture}
    \vspace{-12mm}
    \caption{A fragment of product \ac{fst}.}
    \label{fig:running_example_product_fst_fsa}
\end{figure}

Corollary \ref{corollary:opaque_observations}, as it follows from Theorem \ref{lemma:opaque_runs}, facilitates the computation of the set of opaque observations.
\begin{corollary}
\label{corollary:opaque_observations}
The set of  {opaque observations} is $\opaqueOBS =  L_O(\calA_H) \cap L_O(\calA_H^\dagger).$ 
\end{corollary}

 We compute the intersection product of \ac{nfa}s accepting the languages $L_O(\calA_H)$ and $L_O(\calA^{\dagger}_H)$ in the usual manner as in \cite{baier2008principles}. This results in another \ac{nfa}, which can be determinized into a \ac{dfa} as in \cite{baier2008principles}. We call this \ac{dfa} $\calA^{opaque}=(\hat Q, \Sigma_O, \hat \delta, \hat \init, \hat F)$ as the \emph{opaque-observation \ac{dfa}} as it accepts precisely the set of 
observations that are opaque to P2 (follows from Corollary \ref{corollary:opaque_observations}).

\begin{figure}
   \vspace{1.5mm}
    \centering
    \begin{tikzpicture}[->,>=stealth',shorten >=1pt,auto,node distance=2.5cm,scale=0.7,semithick, transform shape]
    \tikzstyle{every state}=[fill=black!10!white];


    \node[state, initial] (0) at (-3, 3.5) {$(s_\top,q_1)$};
    \node[state] (1) at (-0.5, 3.5) {$(s_1,q_1)$};
    \node[state] (2) at (2, 3.5) {$(s_2,q_1)$};
    \node[state] (3) at (-0.5, 1) {$(s_3,q_1)$};
    \node[state] (4) at (-0.5, -1.25) {$(s_5,q_1)$};
    \node[state] (5) at (-2.5, -1.25) {$(s_6,q_2)$};
    \node[state] (6) at (2, 1) {$(s_\bot,q_1)$};
    \node[state, accepting] (7) at (-4.5, -1.25) {$(s_\bot, q_2)$};


      \path 
        (0) edge   node[sloped]{$\rtimes$} (1)
        (1) edge   node[sloped]{$[s_2, s_3]$} (2)
        (1) edge node[sloped]{$[s_2,s_3]$}(3)
        (1) edge node[pos=0.1,sloped]{$\ltimes$}(6)
        (2) edge [loop right]  node[]{$[s_2, s_3]$} (2)
        (2) edge node [pos=0.2, sloped] {$[s_2, s_3]$} (3)
        (2) edge node[sloped] {$\ltimes$}(6)
        (3) edge [] node [sloped]{$[s_5,s_6]$} (5)
        (3) edge [] node []{$[s_5,s_6]$} (4)
        (3) edge [] node [sloped]{$\ltimes$} (6)
        (3) edge [loop left] node []{$[s_2,s_3]$} (3)
        (4) edge [loop right] node []{$[s_5,s_6]$} (4)
        (4) edge [] node [pos=0.7, sloped]{$\ltimes$} (6)
        (5) edge [loop above] node []{$[s_5,s_6]$} (5)
        (5) edge [] node [sloped]{$\ltimes$} (7)
        (7) edge [loop above] node []{$1$} (7)
        (6) edge [loop below] node []{$1$} (6)

        ;

\end{tikzpicture}
    \vspace{-11mm}
    \caption{A fragment of \ac{nfa} for $L_O(\calA_H)$.}
    \label{fig:running_example_nfa}
    \vspace{-5mm}
\end{figure}
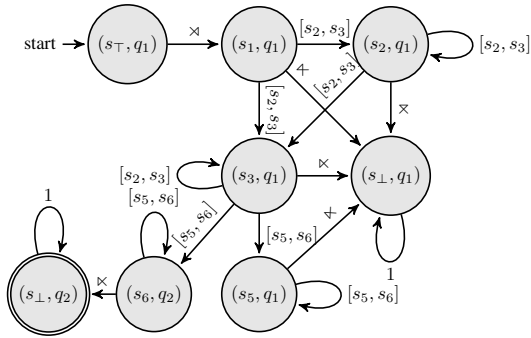

\begin{example}[Part III]
\label{example:part_IV}
      Continuing with Example \ref{example:part_I}, the task specification for P1 is $\psi = \Eventually s_4$ and secret is $\varphi = \Eventually s_6$, represented by the \ac{dfa}s shown in the Fig.~\ref{fig:running_example_dfa}(a) and Fig.~\ref{fig:running_example_dfa}(b). 

    The product \ac{fst} is constructed with the secret \ac{dfa} and the \ac{fst} representation of the observation function. A fragment of the product \ac{fst} is shown in Fig.~\ref{fig:running_example_product_fst_fsa}. Nodes represent the transducer state augmented with the \ac{dfa} state (e.g., $(s_1,q_1)$ where $s_1$ is transducer and $q_1$ is the \ac{dfa} states). The transitions include input and output symbols, like the transducer representation of the observation function.  

  With the product \ac{fst}, we construct an \ac{nfa} accepting $L_O(\calA_H)$ as shown in Fig.~\ref{fig:running_example_nfa}, where the state $(s_\bot,q_2)$ is the final accepting state. Likewise, \ac{nfa} for accepting $L_O(\calA_H^\dagger)$, with $(s_\bot,q_1)$ as the final accepting state. The edges in the \ac{nfa} are labeled with the observations received by P2.
\end{example}

Next, we make use of the above computed opaque-observation \ac{dfa} in our planning algorithm.

\section{Planning Algorithm}
\label{sec:planning_algorithm}

Finally, we can compute a product \ac{mdp} for solving the optimal opacity/transparency planning problem:

\begin{definition}[Product \ac{mdp}] 
\label{def:product_mdp}
Given the \ac{mdp} $M=(S,A, P,\mu_0,s_\top,s_\bot, \calAP, L)$, the task \ac{dfa} $\calA = (Q, \Sigma, \delta, \init, F)$, and the opaque-observation \ac{dfa} $\calA^{opaque} = (\hat Q, \Sigma_O, \hat \delta, \hat \init, \hat F$) with reward, the product \ac{mdp} ($M\times \calA\times \calA^{opaque}$) is a tuple,
\[
\mathcal{M} = (V,A,\mathcal{P},v_0,R_1,R_2)
\]
where 
\begin{itemize}
    \item $V=\{(s,q,\hat{q})\mid s\in S, q\in Q,\hat{q} \in \hat{Q}\}$ is the set of states where each state includes a state from the \ac{mdp} $M$, a state from task \ac{dfa} $\calA$ and a state from the opaque-observations \ac{dfa} $\calA^{opaque}$ respectively. 
    \item $A$ is the set of actions.
    \item $\mathcal{P}: V\times A\rightarrow \dist{V}$ is the probabilistic transition function. For each state $v=(s,q,\hat q) \in V$, an action $a \in A$ enabled from $s$ and state $v'=(s',q',\hat{q}') \in V$, $
    \mathcal{P}(v,a,v')=P(s,a,s')
    $
    if $q' = \delta(q, L(s'))$ and $\hat{q}' = \hat \delta (\hat q, \obs(s,a,s'))$. 
    \item  $v_0 =(s_\top, q_0, \hat q_0)\in V$ is the initial state, where,  $q_0 = \init$ and $\hat q_0 = \hat \init$.
    \item $R_1:V \times A \times V \rightarrow \reals$ is the reward function for the task specification. For a state $(s,q,\hat{q})\in V$, action $a\in A$ and $(s',q',\hat{q}')\in V$, the reward is given by
    \[
    R_1((s,q,\hat{q}), a, (s',q',\hat{q}')) = \begin{cases}
              1, & \text{if}\ q \notin F, q' \in F \\ 
             0, & \text{otherwise}.
             \end{cases}
    \]
    \item $R_2:V\times A\rightarrow\reals$ is the reward function for enforcing opacity. For a state $(s,q,\hat{q})\in V$, action $a\in A$, the reward is given by
        \[
    R_2((s,q,\hat{q}),a) = \begin{cases}
              1, & \text{if}\ \hat{q} \in \hat{F}, a=a_\bot \\ 
             0, & \text{otherwise}.
             \end{cases}
    \]
\end{itemize}
\end{definition}
By construction, for any policy, the accumulated rewards $R_1$ relates to the probability of satisfying the task specification ($\Pr(\mathcal{S}_{\psi}; M^{\pi})$) and the accumulated rewards $R_2$ relates to the probability of enforcing opacity ($ PH(\varphi; M^{\pi}, \obs)$).  Since $M$ specifies that $a_\bot$ is unavailable in $s_\bot$, the agent can only receive the reward once.

We now show that the above product \ac{mdp} can be formulated as a constrained linear program. We formulate the LP problem using the occupancy measures. The optimal opacity enforcement problem (Problem~\ref{prob:problem_statement}) can be solved using the following LP:

\vspace{-5mm} 
\begin{equation}
\label{eq:constrained_lp}
    \begin{aligned}
        & \max_{m} \sum_{v\in V, a\in A} R_2(v, a)m(v,a). \\ 
        \textrm{s.t.} & \sum_{a\in A} m(v,a) = \sum_{v'\in V, a'\in A} \hspace{-3mm}m(v',a')\mathcal{P}(v',a',v) + \nu(v) & \\ \forall v\in V. \\
        & m(v,a) \geq 0. \\
        & \sum_{v\in V, a\in A,v'\in V} \mathcal{P}(v,a,v')R_1(v,a, v')m(v,a) \geq \epsilon.
    \end{aligned}
\end{equation}
where $m(v,a)$ is the occupancy measure that represents the probability of taking an action $a$ from $v$, $\nu$ is the initial distribution of $\mathcal{M}$ such that $\nu(v_0)=1$.  Since the agent can receive the reward  for both $R_1, R_2$ functions only once, for any policy, the total reward (without discounting) is bounded.  Using the solution of LP, the optimal policy for P1 $\pi^{\ast}$ is obtained as $\pi^{\ast}(v,a)=\dfrac{m(v,a)}{\sum_{a'\in A}m(v,a')}$.

\section{Case Study}
\label{sec:case_study}
In this section, we showcase our optimal planning algorithm  on Example \ref{example:part_I} and then in a power plant monitoring scenario. The LPs are solved using the Gurobi solver on an Intel Core i7 CPU @ 3.2 GHz with 32 GB RAM.

\begin{example}[Part IV]
    \label{example:part_VI}
    We now setup the LP formulation for Example \ref{example:part_I}. With the opaque-observations \ac{dfa} $\calA^{opaque}$, \ac{mdp} $M$ and task \ac{dfa} $\calA$, we construct the product \ac{mdp} $\mathcal{M}$. With $\mathcal{M}$, we set up the LP formulations to maximize probabilistic opacity, and maximize probabilistic transparency. 

    \paragraph*{Discussion}
    Table \ref{tab:running_example_opacity_enforcement} and \ref{tab:running_example_opacity_enforcement_transparency} presents results for varying thresholds to maximize opacity, and maximize transparency. Experimental validations of $PH(\varphi)$, $PT(\psi)$, and the probability of task specification are tabulated for P1's policy, obtained by running it for $5000$ instances.
    
    For comparison, a policy that uniformly chooses an action at each state for P1 was applied for $5000$ runs yielded a task satisfaction probability of $0.3402$ and $PH(\varphi)$ of $0.275$.

    Observations from the LP solutions revealed interesting scenarios. In $s_5$, opacity enforcement can be achieved by choosing the terminating action when P2 observes $[s_5, s_6]$. This policy is generated when the threshold is $0.4$. Increasing the threshold prioritizes task satisfaction, leading to a policy requiring P1 to take the actions $b$ and then $a$ to prevent P2 from observing $[s_4]$. Likewise, in $s_7$ with a threshold $0.8$, the policy recommends playing action $a$ with probability $0.786$ and terminate with a probability $0.214$. However, this could result in P2 observing $[s_6]$, thus not enforcing opacity.

\end{example}

\begin{table}
    \vspace{1.2mm}
    \centering
    \begin{tabular}{cccc}
        \textbf{Threshold ($\epsilon$)} & \textbf{Max. $PH(\varphi)$} & \textbf{Exp. $PH(\varphi)$} & \textbf{Exp. Task} \\
        0.4 & 0.7 & 0.6966 & 0.3974\\
        0.6 & 0.6 & 0.6036 & 0.5936\\
        0.8 & 0.4 & 0.4032 & 0.7924\\
    \end{tabular}
    \caption{Results of LP for the running example enforcing opacity.}
    \vspace{-6mm}
    \label{tab:running_example_opacity_enforcement}
\end{table}

\begin{table}
    \centering
    \begin{tabular}{cccc}
        \textbf{Threshold ($\epsilon$)} & \textbf{Max. $PT(\varphi)$} & \textbf{Exp. $PT(\varphi)$} & \textbf{Exp. Task} \\
        0.4 & 0.9828 & 0.9833 & 0.3980\\
        0.6 & 0.9742 & 0.9751 & 0.5966\\
        0.8 & 0.9658 & 0.9667 & 0.7951\\
    \end{tabular}
    \caption{Results for the running example enforcing transparency.}
    \vspace{-10mm}
    \label{tab:running_example_opacity_enforcement_transparency}
\end{table}

\subsection{Opacity enforcement}
\begin{figure}[h]
    \vspace{-3mm}
    \centering    \includegraphics[scale=0.3]{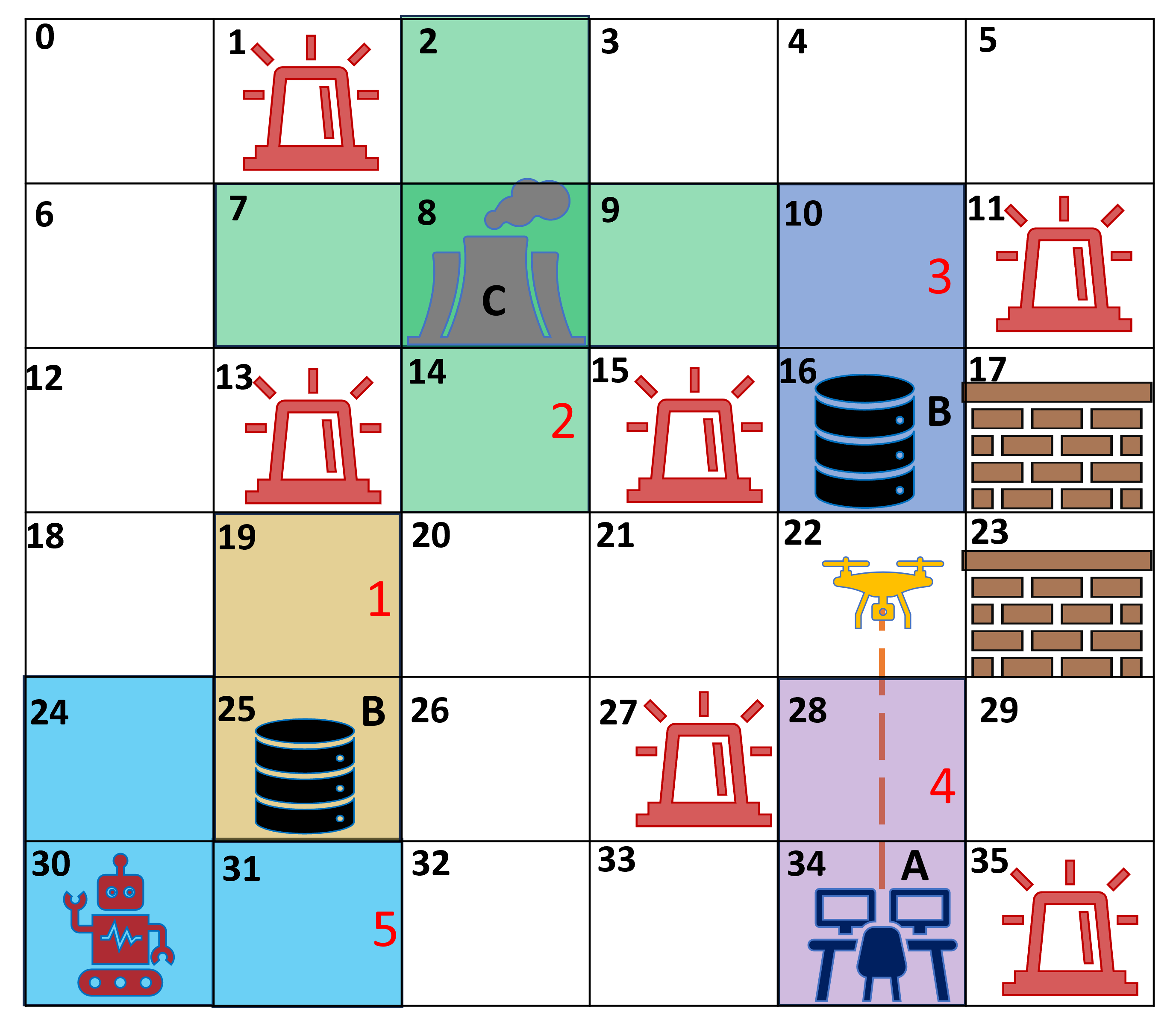}
    \vspace{-2mm}
    \caption{Gridworld depiction of power plant (sensors numbered in red).}
    \vspace{-6mm}
    \label{fig:power_plant_gridworld}
\end{figure}

Consider a power plant as represented in the $6 \times 6$ gridworld shown in Fig.\ref{fig:power_plant_gridworld}. The plant ($\mathbf{C}$) is in cell $8$, a control center ($\mathbf{A}$) in cell $34$ and two data centers ($\mathbf{B}$) in cells $16$ and $25$. Security alarms are placed in the cells $1, 11, 13, 15, 27$ and $35$. A robot performs a routine maintenance task on the plant, specified by the \ac{ltlf} formula $\psi = \Eventually \mathbf{C}$. 

The robot moves in four compass directions. It enters the intended cell with a probability $p=0.6$, and the neighboring cells with probability $(1-p)/2$. For instance, if the robot chooses action $\mathbf{N}$ from cell $30$, it moves to $24$ with $p$ and $30,31$ with $(1-p)/2$. Cells $17$ and $23$ have bouncy walls. If the robot hits the boundary or walls, it stays in its cell. Entering alarmed cells triggers them and disables the robot.


P2 observes the robot through a set of binary range sensors (\textbf{1,2,3,4}, yellow, green, indigo, and purple resply.) and a precision sensor (\textbf{5}, blue). The binary sensors return a value $\mathsf{True}$ when the robot is in the range and $\mathsf{False}$ otherwise, while the precision sensors return the exact position of the robot when the robot is in range. The coverage of each of the sensors is shown in Fig. \ref{fig:power_plant_gridworld}. P2 also deploys as a dynamic sensor flying between cells $34$ and $16$, with a $0.65$ probability to move to the next cell and $0.35$ to stay. It changes direction following the dynamics when at cells $22$ or $34$. The drone's camera provides precision sensing for the drone's cell and the north cell (e.g., at $34$, it precisely monitors $34$ and $28$).   


We consider the scenario where the robot, controlled by P1, aims to enforce opacity under task constraints. P1 has the task $\psi = \Eventually \mathbf{C}$. Additionally, P1 must perform a secret task 
specified by the \ac{ltlf} formula $\varphi = \Eventually \mathbf{B} \land \Eventually \mathbf{A}$. 
The opaque-observations \ac{dfa}, with $585$  states, is constructed in an average time of $90.1576s$ for the considered configuration. Subsequently, the product \ac{mdp}, with $7596$ states is constructed in $2.9703s$. The LP for optimal policy is solved in an average time of $3099.1386s$.  

Table \ref{tab:opacity_enforcement_power_plant} summarizes opacity enforcement probabilities at various task satisfaction thresholds. 
As the task specification threshold increases,  the probability of enforcing opacity decreases under the solved policy.  The policy suggests terminating the game after satisfying the secret for lower thresholds. With a higher threshold, the robot must take the specific path through $16$, $10$, and $9$ or $4$ to reach the target in $8$. Notably, taking the path to $16$ from $22$ reveals to P2 that $16$ has been reached, compromising opacity. Thus, opacity is only enforced when reaching $34$ (or $\mathbf{A}$) is concealed.

To assess the impact of sensors on probabilistic opacity, we conduct experiments under the following conditions: \begin{inparaenum}
    \item No drone, 
    \item The drone flies between the cells $34$, $22$ and $25$.
    \item No sensor $\mathbf{3}$. 
\end{inparaenum} The results are tabulated in Table \ref{tab:opacity_enforcement_power_plant_comparison_exp}, specifically for a threshold of $0.4$ (due to page limitations). From Table \ref{tab:opacity_enforcement_power_plant_comparison_exp}, it is evident that in the absence drone or sensor $\mathbf{3}$, P1 achieves better opacity enforcement. As P2's sensing capabilities decrease, the probability of opacity enforcement increases as expected. 

\begin{table}
    \vspace{1.2mm}
    \centering
    \begin{tabular}{ccc}
        \textbf{Experiment} & \textbf{Prob. Opacity} & \textbf{Exp. Opacity}\\
         No drone & $0.7058$ & $0.7102$\\
        Drone covers larger area & $0.6154$ & $0.6179$\\
        No Sensor $\mathbf{3}$ &$0.7950$ & $0.7894$ \\
    \end{tabular}
    \caption{Comparison of opacity enforcement (with threshold $0.4$).}
    \vspace{-7mm}\label{tab:opacity_enforcement_power_plant_comparison_exp}
\end{table}



\begin{table}
    \centering
    \begin{tabular}{cccc}
        \textbf{Threshold ($\epsilon$)} & \textbf{Prob. Opacity} & \textbf{Exp. Opacity} & \textbf{Exp. Task}\\
         0.4& 0.6766 & 0.6678 & 0.4014\\
         0.6&  0.5333& 0.5401 & 0.6021\\
         0.7&  0.4611& 0.4625 & 0.7056\\
         0.8& 0.3448& 0.3359 & 0.8046\\
         0.95& 0.1079& 0.1085 & 0.9504\\
    \end{tabular}
    \caption{Enforced probabilistic opacity and experimental validation. }
    \vspace{-7mm}\label{tab:opacity_enforcement_power_plant}
\end{table}

\subsection{Transparency enforcement}
\begin{figure}
   \vspace{1mm}
    \centering
    \includegraphics[scale=0.3]{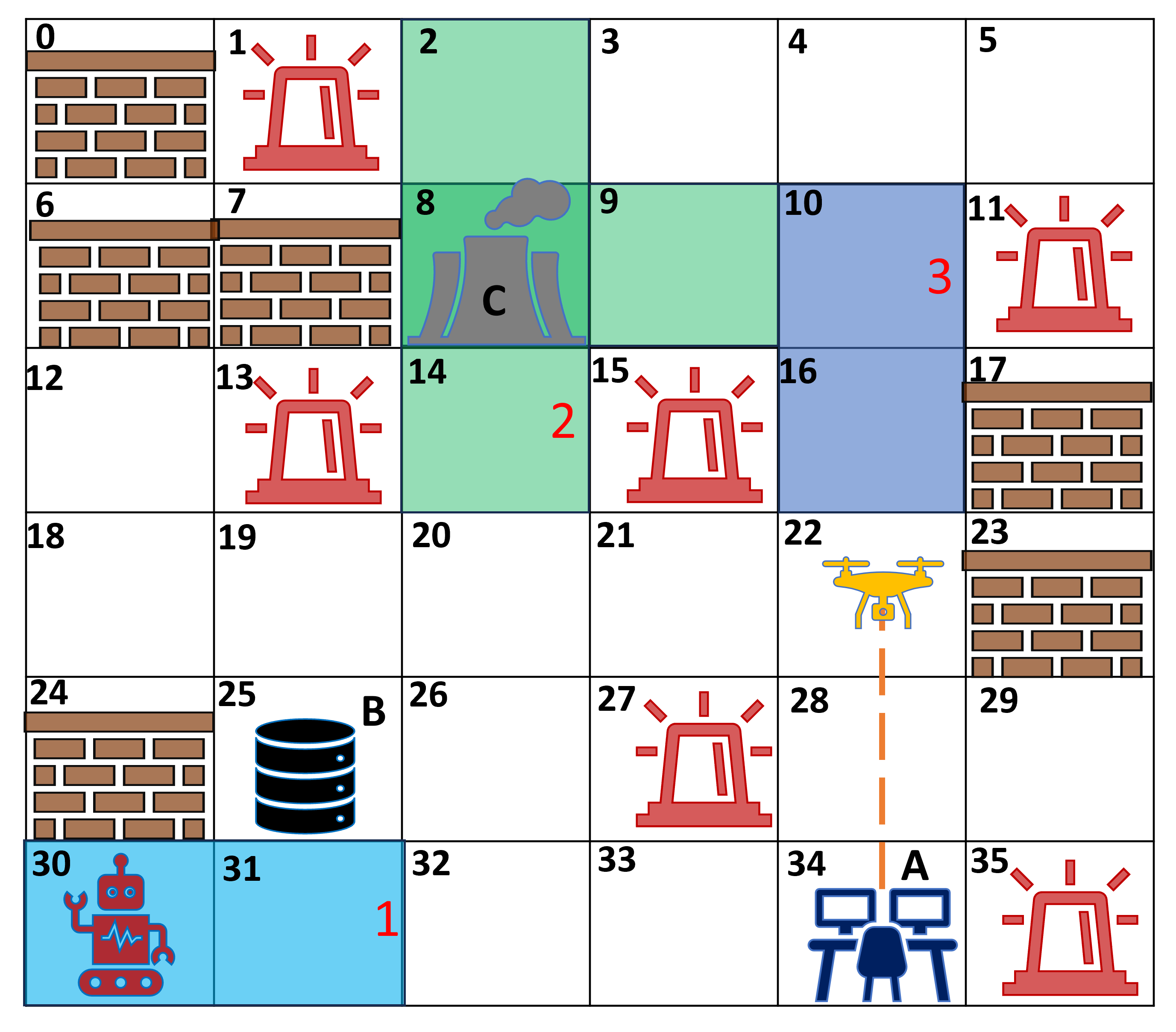}
    \vspace{-3mm}
    \caption{Gridworld depiction for enforcing transparency.}
    \vspace{-3mm}
    \label{fig:gridworld_representation_transparency}
\end{figure}

Consider a case where P1 enforces transparency, in a setup similar to previous case, shown in the Fig. \ref{fig:gridworld_representation_transparency}. The environment includes one data center ($\mathbf{B}$) in $25$ and a control center ($\mathbf{A}$) in $34$. Three sensors are present: sensor $\mathbf{1}$ a static precision sensor that monitors $30$ and $31$, sensor $\mathbf{2}$ and sensor $\mathbf{3}$, static binary sensors that monitor $2, 8, 9, 14$, and $10, 16$ resply. Additionally, the drone operates between $34$ and $22$ with the same dynamics as before.

The robot maintains the previous task specification $\psi = \Eventually \mathbf{C}$, and P1's specification for transparency on P2 is $\varphi = \Eventually \mathbf{B} \land \Eventually \mathbf{A}$.

Table \ref{tab:transparency_enforcement_power_plant} summarizes transparency enforcement probabilities for various task specification thresholds and presents the results for P1 following the optimal policy.

\begin{table}
    \centering
    \begin{tabular}{cccc}
        \textbf{Threshold ($\epsilon$)} & \textbf{Prob. Transp.} & \textbf{Exp. Transp.} & \textbf{Exp. Task}\\
         0.4& 0.8795 & 0.8784 & 0.4022 \\
         0.6&  0.6802& 0.6814& 0.5999 \\
         0.7&  0.5581& 0.5623& 0.7012\\
         0.8& 0.4100& 0.4105& 0.8064\\
         0.95& 0.1204& 0.1198& 0.9503 \\
    \end{tabular}
    \caption{Enforced probabilistic transparency and validation. }
    \vspace{-10mm}
    \label{tab:transparency_enforcement_power_plant}
\end{table}

\section{Conclusion}
\label{sec:conclusion}
We introduced constrained probability planning methods for \ac{mdp}s to optimize opacity/transparency under task constraints. Our approach starts with building  computation models based on \ac{fst} to capture the observation function of the observer that maps the input languages --- state trajectories  into the output languages --- observed trajectories. Then, by performing the product operations between the \ac{fst} and the \ac{dfa} accepting the secret, we can derive another \ac{dfa}
 that accepts all possible observations that enforce the opacity of the secret to the observer. 
 Thus, we can formulate the probabilistic planning with opacity as a constrained \ac{mdp}, augmented with the task state and the state in the \ac{dfa} of the opaque observations. The dual problem of transparency can be solved by replacing the maximization with a minimization in the objective function for the constrained \ac{mdp}.
 Through experimental analysis, we investigated the impact of sensors and sensor configurations on opacity and transparency enforcement.
 The construction of opaque/transparent observations can  be extended for optimizing opacity in partially observable systems  or games with  partial observations, involving both collaborative and competitive interactions.


\bibliography{refs}

\begin{thebibliography}{10}

\bibitem{baier2008principles}
C.~Baier and J.-P. Katoen.
\newblock {\em Principles of Model Checking}.
\newblock MIT press, 2008.

\bibitem{berard2015probabilistic}
B.~B{\'e}rard, K.~Chatterjee, and N.~Sznajder.
\newblock Probabilistic opacity for markov decision processes.
\newblock {\em Information Processing Letters}, 115(1):52--59, 2015.

\bibitem{berard2015quantifying}
B.~B{\'e}rard, J.~Mullins, and M.~Sassolas.
\newblock Quantifying opacity.
\newblock {\em Mathematical Structures in Computer Science}, 25(2):361--403, 2015.

\bibitem{bryans2008opacity}
J.~W. Bryans, M.~Koutny, L.~Mazar{\'e}, and P.~Y. Ryan.
\newblock Opacity generalised to transition systems.
\newblock {\em International Journal of Information Security}, 7:421--435, 2008.

\bibitem{bryans2005modelling}
J.~W. Bryans, M.~Koutny, and P.~Y. Ryan.
\newblock Modelling opacity using petri nets.
\newblock {\em Electronic Notes in Theoretical Computer Science}, 121:101--115, 2005.

\bibitem{cassez2012synthesis}
F.~Cassez, J.~Dubreil, and H.~Marchand.
\newblock Synthesis of opaque systems with static and dynamic masks.
\newblock {\em Formal Methods in System Design}, 40:88--115, 2012.

\bibitem{de2013linear}
G.~De~Giacomo and M.~Y. Vardi.
\newblock Linear temporal logic and linear dynamic logic on finite traces.
\newblock In {\em Proceedings of the Twenty-Third international joint conference on Artificial Intelligence}, pages 854--860. ACM, 2013.

\bibitem{jacob2016overview}
R.~Jacob, J.-J. Lesage, and J.-M. Faure.
\newblock Overview of discrete event systems opacity: Models, validation, and quantification.
\newblock {\em Annual reviews in control}, 41:135--146, 2016.

\bibitem{keroglou2018probabilistic}
C.~Keroglou and C.~N. Hadjicostis.
\newblock Probabilistic system opacity in discrete event systems.
\newblock {\em Discrete Event Dynamic Systems}, 28:289--314, 2018.

\bibitem{lin2011opacity}
F.~Lin.
\newblock Opacity of discrete event systems and its applications.
\newblock {\em Automatica}, 47(3):496--503, 2011.

\bibitem{liu2024approximate}
S.~Liu, X.~Yin, D.~V. Dimarogonas, and M.~Zamani.
\newblock On approximate opacity of stochastic control systems.
\newblock {\em arXiv preprint arXiv:2401.01972}, 2024.

\bibitem{maubert2011opacity}
B.~Maubert, S.~Pinchinat, and L.~Bozzelli.
\newblock Opacity issues in games with imperfect information.
\newblock In {\em In 2nd Int. Symp. on Games, Automata, Logics and Formal Verification}, pages 87--101, 2011.

\bibitem{mazare2004using}
L.~Mazar{\'e}.
\newblock Using unification for opacity properties.
\newblock {\em Proceedings of the 4th IFIP WG1}, 7:165--176, 2004.

\bibitem{saboori2010verification}
A.~Saboori.
\newblock {\em Verification and Enforcement of State-Based Notions of Opacity in Discrete Event Systems}.
\newblock University of Illinois at Urbana-Champaign, 2010.

\bibitem{saboori2007notions}
A.~Saboori and C.~N. Hadjicostis.
\newblock Notions of security and opacity in discrete event systems.
\newblock In {\em IEEE Conference on Decision and Control}, pages 5056--5061, 2007.

\bibitem{saboori2008verification}
A.~Saboori and C.~N. Hadjicostis.
\newblock Verification of initial-state opacity in security applications of des.
\newblock In {\em 9th International Workshop on Discrete Event Systems}, pages 328--333. IEEE, 2008.

\bibitem{saboori2011coverage}
A.~Saboori and C.~N. Hadjicostis.
\newblock Coverage analysis of mobile agent trajectory via state-based opacity formulations.
\newblock {\em Control Engineering Practice}, 19(9):967--977, 2011.

\bibitem{saboori2012}
A.~Saboori and C.~N. Hadjicostis.
\newblock Opacity-enforcing supervisory strategies via state estimator constructions.
\newblock {\em IEEE Transactions on Automatic Control}, 57(5):1155--1165, 2012.

\bibitem{saboori2014current}
A.~Saboori and C.~N. Hadjicostis.
\newblock Current-state opacity formulations in probabilistic finite automata.
\newblock {\em IEEE Transactions on Automatic Control}, 59(1):120--133, 2014.

\bibitem{udupa2023opacity}
S.~Udupa, H.~Rahmani, and J.~Fu.
\newblock Opacity-enforcing active perception and control against eavesdropping attacks.
\newblock In {\em International Conference on Decision and Game Theory for Security}, pages 329--348. Springer, 2023.

\bibitem{wu2013comparative}
Y.-C. Wu and S.~Lafortune.
\newblock Comparative analysis of related notions of opacity in centralized and coordinated architectures.
\newblock {\em Discrete Event Dynamic Systems}, 23(3):307--339, 2013.

\bibitem{wu2014synthesis}
Y.-C. Wu and S.~Lafortune.
\newblock Synthesis of insertion functions for enforcement of opacity security properties.
\newblock {\em Automatica}, 50(5):1336--1348, 2014.

\bibitem{xie2021opacity}
Y.~Xie, X.~Yin, and S.~Li.
\newblock Opacity enforcing supervisory control using nondeterministic supervisors.
\newblock {\em IEEE Transactions on Automatic Control}, 67(12):6567--6582, 2021.

\bibitem{yao2022sensor}
J.~Yao, X.~Yin, and S.~Li.
\newblock Sensor deception attacks against initial-state privacy in supervisory control systems.
\newblock In {\em 2022 IEEE 61st Conference on Decision and Control}, pages 4839--4845. IEEE, 2022.

\bibitem{yin2019infinite}
X.~Yin, Z.~Li, W.~Wang, and S.~Li.
\newblock Infinite-step opacity and k-step opacity of stochastic discrete-event systems.
\newblock {\em Automatica}, 99:266--274, 2019.

\bibitem{yingrui2023}
Y.~Zhou, Z.~Chen, and Z.~Liu.
\newblock Verification and enforcement of current-state opacity based on a state space approach.
\newblock {\em European Journal of Control}, 71:100795, 2023.

\end{thebibliography}
\bibliographystyle{abbrv}
\end{document}